\documentstyle[prb,aps,amssymb,multicol,epsf,psfig,pstricks]{revtex}

\newcommand{\onecolm}{
  \end{multicols}
  \vspace{-3.5ex}
  \noindent\rule{0.5\textwidth}{0.1ex}\rule{0.1ex}{2ex}\hfill
}
\newcommand{\twocolm}{
  \hfill\raisebox{-1.9ex}{\rule{0.1ex}{2ex}}\rule{0.5\textwidth}{0.1ex}
  \vspace{-4ex}
  \begin{multicols}{2}
}

\newcommand{\beq}{\begin{equation}}
\newcommand{\beqa}{\begin{eqnarray}}
\newcommand{\eeq}{\end{equation}}
\newcommand{\eeqa}{\end{eqnarray}}
\newcommand{\nbeqa}{\begin{eqnarray*}}
\newcommand{\neeqa}{\end{eqnarray*}}
\newenvironment{neqblock}[1]{\[\begin{array}{#1}}{\end{array}\]}

\newenvironment{eqblock}[2]{\beq\label{#2}\begin{array}{#1}}{\end{array}
                                \eeq}
\newcommand{\beqb}{\begin{eqblock}}
\newcommand{\eeqb}{\end{eqblock}} 
\newcommand{\nbeqb}{\begin{neqblock}}
\newcommand{\neeqb}{\end{neqblock}} 
\newcommand{\sDelta}{{\scriptstyle\Delta}}
\newcommand{\bra}[1]{\left\langle #1 \right |}
\newcommand{\ket}[1]{\left | #1 \right\rangle}

\def\CC{{\rm\kern.24em \vrule width.04em height1.46ex depth-.07ex
\kern-.30em C}}
\def\P{{\rm I\kern-.25em P}}
\def\RR{{\rm
         \vrule width.04em height1.58ex depth-.0ex
         \kern-.04em R}}
\def\id{{\rm 1\kern-.22em l}}

\begin{document}

\title{Exact solution of generalized Schulz-Shastry type models}
        
\author{Andreas Osterloh$^{1,3}$, Luigi Amico$^{2}$, and 
Ulrich Eckern$^{1}$}
\address{$^{1}$ Institut f\"ur Physik,Universit\"at Augsburg, D-86135 Augsburg.}
\address{$^{2}$ Dipartimento di Metodologie Fisiche e Chimiche
per l'Ingegneria, Facolt\'a di Ingegneria,
Universit\'a di Catania \& INFM, viale A. Doria 6, I-95129 Catania.}
\address{$^{3}$ andreas.osterloh@physik.uni-augsburg.de}

\maketitle

\bigskip

\begin{abstract}

A class of integrable one-dimensional models presented by Shastry and Schulz 
is consequently extended to the whole class of one-dimensional Hubbard- 
or XXZ-type models with correlated gauge-like hopping.
A complete characterization concerning solvability by coordinate Bethe
ansatz of this class of models is found.
\medskip


\end{abstract}

\begin{multicols}{2}

\section{Introduction}

Integrability has been considered to be one of 
the most striking properties of a model for some time. 
However, the interest 
was confined to an abstract level since integrable quantum systems 
typically live in one spatial dimension
(but they can be mapped onto classical $1+1$ dimensional 
systems~\cite{TAKHFADDEEV,ZAMOLO}). In the meantime,
one-dimensional systems have become 
experimental reality (Quantum Hall bars~\cite{QHALL}, polymers~\cite{POLY}, 
charge density wave systems~\cite{CDW}) and thus
integrable models have gained  immediate relevance for real physical problems.
The Coordinate Bethe Ansatz (CBA)~\cite{BETHE,YANGYANG} solvability is 
the first milestone in proving integrability.
The (Quantum) Inverse Scattering Method together with the algebraic 
Bethe Ansatz complete the CBA procedure to 
construct integrable models\cite{TAKHFADDEEV,KOREPIN}. 
However, the interrelation between solvability by algebraic Bethe
ansatz and CBA is still controversial. 
But solvability by algebraic Bethe Ansatz is a strong hint for
solvability by CBA within an appropriately chosen basis.
In the case of the quantum impurity problems, for instance, the standard CBA procedure 
with plane waves fails, in contrast to using a basis consisting of
solitary waves~\cite{ZAMOLO,PROPERBASIS}.
The reason for this is that the bulk theory already describes fully interacting 
electrons and there is no (free) single particle description of the system.
As a result the scattering with the impurity is ``diffractive'' in the plane-wave basis.  
Instead, it turns out to be factorizable if the many-electron 
wave function is written in terms of kinks and anti--kinks.
\\
A breakthrough in the theory of integrable models was
the solution of the Hubbard Model (HM) obtained by E. H. Lieb and F. Y. Wu 
\cite{LIEBWU}. The HM is a model including
on-site Coulomb interaction for electrons moving in an atomic lattice. It is
believed to capture important features of high-$T_C$ superconductivity
\cite{HM-SC} and has a metal-insulator transition.

In a recent paper, H. J. Schulz and B. S. Shastry found a new class of 
solvable  one-dimensional Hubbard and XXZ type models\cite{SCHULZ}. 
The modification of the original HM and XXZ model consisted in a 
configuration dependent unitary factor in the hopping term. 
This can be interpreted as 
an interaction of the charged particles with a gauge-field,
generated by the density of particles. 
The structure of the unitary factor was
$\exp [{\rm i}\hat{N}]$, where $\hat{N}$ is a mono-linear functional of particle-number
operators; we  term such models ``single particle correlated hopping ($1$-CH)'' models.
\\
The idea behind Schulz' and Shastry's approach is finding 
a basis (through a unitary transformation of the original Fock basis) in  which the model 
takes the form of the original Hubbard or XXZ model up to boundary twists which do not 
affect their solvability~\cite{SUTHERLAND}.
We point out that this is equivalent to equipping the plane waves entering the CBA 
with phase factors canceling exactly the configuration dependent gauge fields 
in the hopping term. 
In the present paper we will generalize such an idea  to 
consider hopping in which $\hat{N}$ is a multilinear functional of 
particle-number operators. We shall call the resulting models $n$--CH models.  
We will answer the question which $n$--CH-Hubbard/XXZ 
models can be mapped unitarily onto a corresponding uncorrelated but
twisted model. This finally proves solvability of the model. 
Conversely, a non-removable correlated hopping
destroys solvability by CBA, since the $S$ matrix becomes 
con\-figu\-ra\-tion dependent (see Appendix \ref{SCONFIG}).

The paper is laid out as follows. In section \ref{1CH}, we will study 
the most general form of $1$--CH HMs, hence including Shastry--Schulz models. 
Most of the features of the general problem already occur here.
Section \ref{MC-models} accounts for conserved quantum numbers other than the spin
orientation of electrons. 
Multi-chain models where the chains interact with each other 
exclusively gauge-like, will be considered.
The central results obtained in sections \ref{1CH} and \ref{MC-models} will be used to
discuss higher correlated hopping in section \ref{2CH}. The $2$-CH will be
treated in detail, enlightening the approach to general $n$-CH.
Finally, conclusions will be drawn in section \ref{CONCL}.

A special class of unitary transformations was used to remove the
gauge-like correlation terms from the hopping. In Appendix 
\ref{GENERAL-UNITARY} the effect of 
the complementary class of unitary transformations will be elaborated,
always restricting on automorphic mappings on the class of 
twisted $n$-CH models of Hubbard- or XXZ-kind.
Two propositions needed in this section are proven in 
Appendix \ref{QED} (see also Ref. \onlinecite{EXTERN}).

\section{Single-particle correlated hopping}\label{1CH}
In this section we discuss a simple generalization
of Schulz-Shastry models.
Such models have the following Hubbard-type Hamiltonian
\begin{eqnarray}
\label{SS}
H &=& - t\, \sum_{j,\sigma } 
\, \biggl\{ c_{j+1,\sigma}^\dagger c_{j,\sigma} \exp({\rm i}\gamma_j(\sigma))\times \nonumber \\
&&\ \times\exp\Bigl[{\rm i}\sum_{l}^{}\bigl(\alpha_{j,l}(\sigma)n_{l,-\sigma}
   +  A_{j,l}(\sigma)n_{l,\sigma}\bigr)\Bigr] + {\rm h.c.} \biggr\} + \nonumber \\ 
 && \qquad + V\, \sum_{ i}n_{{i},\uparrow} n_{{ i},\downarrow}. 
\end{eqnarray}
where $\{c_{j,\sigma}, c_{l,\sigma'}^\dagger \}=\delta_{\sigma,\sigma'} \delta_{j,l}$, 
$\{c_{j,\sigma}, c_{l,\sigma'} \}=0$, and $n_{l,\sigma}:= c_{l,\sigma}^\dagger c_{l,\sigma}$.
The parameters $t$, $V$ are the hopping amplitude and the Coulomb repulsion respectively.
The class of models~(\ref{SS}) is a generalization of Schulz--Shastry 
models, since a) the parameter $A$ occurs, which means correlation between particles
with the same spin orientation, and b) the spin- and coordinate dependence of
$A \;, \alpha \;, \gamma$ is unrestricted here~\cite{NOTESS}.
\\
We point out that $A_{j,j+1}$ can be 
set to an arbitrary value for all $j$, without affecting the physics of the model 
(since $n_{l,\sigma}\in \{0,1\}\ \Rightarrow\ 
c_{j+1,\sigma}^\dagger n_{j+1,\sigma}\equiv 0 $).
We name parameters like $A_{j,j+1}$ ``irrelevant''.
A similar argument
holds for the parameter $A_{j,j}$: contributions from $n_{j,\sigma}$
arise only if $n_{j,\sigma}=1$ because of
$c_{j,\sigma}n_{j,\sigma}\equiv c_{j,\sigma}$. 
Hence, this term can be included in the parameter $\gamma_j(\sigma)$.
Parameters like
$A_{j,j}$ will be called ``subrelevant'' throughout this paper.
Irrelevant as well as subrelevant parameters appear as soon as 
phase--correlations among particles having the same 
spin orientation as the hopping particle are involved. 
\\
It is worthwhile noting that Hamiltonian (\ref{SS}) 
is not diagonalizable by {\em direct} CBA 
since the scattering matrix is configuration dependent 
(see Appendix \ref{SCONFIG}).
This destroys the factorizability of a many-particle $S$ matrix into 
two-particle $S$ matrices. Thus, we first remove the phases in the
hopping term of (\ref{SS}) by a unitary transformation and then, we
can diagonalize the transformed Hamiltonian by CBA. 
\\
The ansatz for the unitary transformation is achieved through the
operator
\begin{equation}\label{U}
U:=\exp\bigl[ {\rm i}(\xi^{\mu,\nu}_{l,m}n_{l,\mu}n_{m,\nu}+\zeta_{l,\mu}n_{l,\mu})
\bigr]=: \exp({\rm i} S),
\end{equation}
(we use the sum convention)  where $ \xi_{i,j},\ \zeta_{l,m}\in \RR$ are 
unknown variables which 
have to be fixed for cancelling the unitary prefactor in the hopping term
of~(\ref{SS}).
Since an antisymmetric part in the parameter $\xi_{i,j}$ 
vanishes after summation, it can be defined fully 
symmetric: $\xi^{\mu,\nu}_{l,m}=\xi^{\nu,\mu}_{m,l}$.
We can further choose $\xi^{\mu,\mu}_{m,m}=0$ (a non-zero $\xi^{\mu,\mu}_{m,m}$ 
can be included in the parameter $\zeta_{l,m}$).
\\
We locally transform the Hamiltonian by $U$:  
$
c_{j,\sigma} \stackrel{U}{\longrightarrow} U\, c_{j,\sigma}\, U^{\! -1} .
$
Number operators remain unchanged, but the hopping term is altered
\begin{eqnarray*}
c_{j+1,\sigma}^\dagger c_{j,\sigma} &\stackrel{U}{\longrightarrow}&
c_{j+1,\sigma}^\dagger c_{j,\sigma}
\exp\Bigl[2{\rm i}\bigl(\xi^{\sigma,\mu}_{j+1,m}-\xi^{\sigma,\mu}_{j,m}\bigr)
n_{m,\mu}\Bigr]\\
&& \ \exp\Bigl[{\rm i}\bigl(\zeta_{j+1,\sigma}-\zeta_{j,\sigma}-2\xi^{\sigma,\sigma}_{j,j+1}
\bigr)
\Bigr].
\end{eqnarray*}
The unitary factors in 
the hopping term of (\ref{SS}) are removed if 
\begin{eqnarray}
2(\xi^{\sigma,-\sigma}_{j,m}-\xi^{\sigma,-\sigma}_{j+1,m}) 
&\stackrel{!}{=}& \alpha_{j,m}(\sigma)  \label{ZERO1}\\
(\zeta_{j,\sigma}-\zeta_{j+1,\sigma}+2\xi^{\sigma,\sigma}_{j,j+1})
&\stackrel{!}{=}& \gamma_{j}(\sigma), \label{ZERO3}
\end{eqnarray}
and
\begin{equation}
\label{ZERO2}
2(\xi^{\sigma,\sigma}_{j,m}-\xi^{\sigma,\sigma}_{j+1,m}) 
\stackrel{!}{=} A_{j,m}(\sigma)\ ;\  m\in\{1,\dots , L \} \setminus
\{j,j+1 \}.
\end{equation}
For periodic boundary conditions (PBC), Eqs. (\ref{ZERO1})--(\ref{ZERO2})
for $j=L$ represent the jump across the boundary.
Admitting boundary phases in company with the boundary's crossing,
Eqs.~(\ref{ZERO1})--(\ref{ZERO2}) are modified as 
\begin{eqnarray}
2(\xi^{\sigma,-\sigma}_{L,m}-\xi^{\sigma,-\sigma}_{1,m})  
&\stackrel{!}{=}& \alpha_{L,m}(\sigma)-\phi^{(1)}_{\uparrow\downarrow}(\sigma),
\label{BZERO1}\\
(\zeta_{L,\sigma}-\zeta_{1,\sigma}+2\xi^{\sigma,\sigma}_{L,1})
&\stackrel{!}{=}& \gamma_{L}(\sigma)-\phi(\sigma) . 
\label{BZERO3}
\end{eqnarray}
and
\begin{equation}
\label{BZERO2}
2(\xi^{\sigma,\sigma}_{L,m}-\xi^{\sigma,\sigma}_{1,m}) 
\stackrel{!}{=} A_{L,m}(\sigma)-\phi^{(1)}_{\uparrow\uparrow}(\sigma) 
\ ;\  m\in\{2,\dots , L-1 \} , 
\end{equation}
In fact, the relations~(\ref{ZERO1})--(\ref{ZERO2}) and~(\ref{BZERO1})--(\ref{BZERO2})
constitute a system of recursive relations for 
$\xi^{\sigma,\sigma'}_{j,m}$ and $\zeta_{j,\sigma}$.
\\
We will now discuss the exclusions in Eq. (\ref{ZERO2}).
The corresponding part of the transformed hopping term for $m=j$ is
(since $\xi^{\sigma,\sigma}_{j,j}=0$)
$
c_{j+1,\sigma}^\dagger c_{j,\sigma}
\exp\Bigl[{\rm i}\bigl(2\xi^{\sigma,\sigma}_{j+1,j}
+ A_{j,j}(\sigma)\bigr)
n_{j,\sigma}\Bigr].
$
This term is non-zero only if $n_{j,\sigma}=1$. Hence, this ``correlation'' factor
is equivalent to 
$
c_{j+1,\sigma}^\dagger c_{j,\sigma}
\exp\Bigl[{\rm i}\bigl(2\xi^{\sigma,\sigma}_{j+1,j}+A_{j,j}(\sigma)\bigr)\Bigr].
$
For $m=j+1$ it is
$
c_{j+1,\sigma}^\dagger c_{j,\sigma}
\exp\Bigl[-{\rm i}\bigl(
2\xi^{\sigma,\sigma}_{j,j+1}-
A_{j,j+1}(\sigma)\bigr)
n_{j+1,\sigma}\Bigr].
$
This term is non-zero only if $n_{j+1,\sigma}=0$. Hence, this ``correlation'' factor
is equivalent to
$
c_{j+1,\sigma}^\dagger c_{j,\sigma}
$
irrespective of what value
$A_{j,j+1}(\sigma)-2\xi^{\sigma,\sigma}_{j,j+1}$ may take. 
As a consequence, there is no condition for $m=j+1$ in Eq. (\ref{ZERO2}), and
$A_{j,j}(\sigma)+2\xi^{\sigma,\sigma}_{j,j+1}$ enters as a modification 
of Eq. (\ref{ZERO3}) for $\gamma_j$ (see Appendix \ref{DETAILS}).
\begin{figure}
\begin{minipage}{80mm}
\begin{minipage}[h]{80mm}
\psfig{width=80mm,height=80mm,file=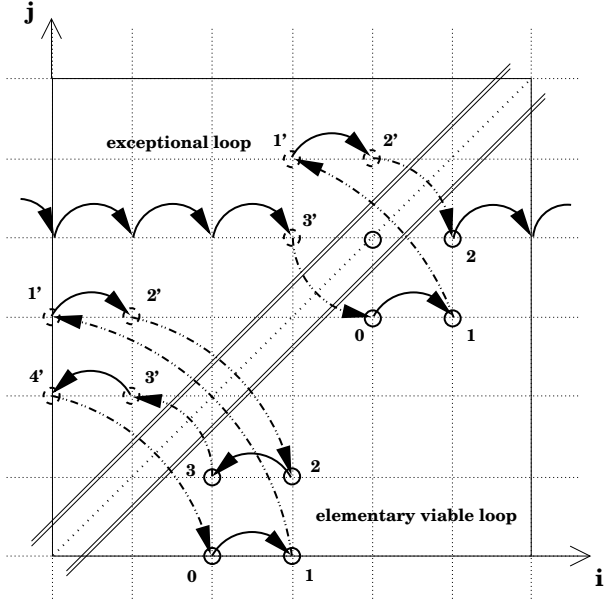}
\end{minipage}
\begin{center}
\begin{minipage}[h]{78mm}
\caption{
\label{criticalloop}
Elementary viable (leading to (\ref{INT2})) and 
not elementarily viable (called ``exceptional'') loops, are drawn 
in the space of indices 
$(i,j)$ of $\xi^{\sigma,\sigma}_{i,j}$ (the crossings of the dotted
lines are the possible index points). The 
expceptional loop  determines the boundary phase for the parameter $A$.
Full arrows mean an application of the recursive relation (\ref{ZERO2}),
which can transport in horizontal direction only.
The other arrows visualize the application of the symmetry $\xi_{i,j}=\xi_{j,i}$. 
The double lines are reminders of the missing connection between 
$\xi_{j-1,j}^{\sigma,\sigma}$ and $\xi_{j,j}^{\sigma,\sigma}$. Hence,
it must not be crossed by the full arrows.}
\end{minipage}
\end{center}
\end{minipage}
\end{figure}
The system of equations~(\ref{ZERO1})--(\ref{ZERO3})  
cannot be solved for arbitrary $\alpha$ and $A$ 
(the number of parameters 
on the right-hand side is larger than on the left-hand side).
Together with the symmetry of $\xi_{i,j}$, Eqs. (\ref{ZERO1}) and
(\ref{ZERO2}) define the effect of an increase in both site  
indices of $\xi_{i,j}$; namely an increase of $\xi$ by the related
$\alpha$ respectively $A$. 
Starting from an initial parameter, say $\xi_{1,1}:=0$, every $\xi_{i,j}$
is then defined by passing from $(1,1)$ to $(i,j)$ and summing up the
contributions from the recursive relation. 
$\xi_{i,j}$ is well defined if this procedure is path independent.
This is equivalent to demanding that contributions from closed loops
in the $(i,j)$ plane vanish.
To verify this, it is sufficient facing the smallest possible loops:
$(j,m) \rightarrow (j+1,m) \rightarrow (j+1,m+1)\rightarrow (j,m+1) \rightarrow (j,m)$
(which we will henceforth call ``elementary''). 
Applying Eq. (\ref{ZERO1}) respectively Eq. (\ref{ZERO2}), we obtain
\begin{equation}
\alpha_{j,m+1}(\sigma)
-\alpha_{j,m}(\sigma) = \alpha_{m,j+1}(-\sigma) - \alpha_{m,j}(-\sigma)\, , \label{INT1} 
\end{equation}
and
\begin{equation}
A_{j,m+1}(\sigma)- A_{j,m}(\sigma) =
A_{m,j+1}(\sigma) - A_{m,j}(\sigma) \label{INT2}
\end{equation}
for $m\neq j,j\pm 1$.
We call the conditions (\ref{INT1}), (\ref{INT2})) ``closedness conditions''.
The recursive relation and the closedness condition can be written in
a more compact and clear form in terms of the discrete gradient,
defined by $\partial_x f(x) := f(x+1) -f(x)$. The recursive relations
(\ref{ZERO1}) and (\ref{ZERO2}) then read
\begin{eqnarray}
-2\partial_x \xi^{\sigma,-\sigma}(x,y) \equiv -2\partial_x
\xi^{-\sigma,\sigma}(y,x) &=& \alpha(x,y;\sigma) \\
-2\partial_x \xi^{\sigma,\sigma}(x,y) \equiv -2\partial_x
\xi^{\sigma,\sigma}(y,x) &=& A(x,y;\sigma) 
\end{eqnarray}
and the closedness conditions take the form
\begin{eqnarray}
\partial_y \alpha(x,y;\sigma) &=& \partial_x \alpha(y,x;-\sigma) \\
\partial_y A(x,y;\sigma) &=& \partial_x A(y,x;\sigma)\ ;\ x\neq y.
\end{eqnarray}
With these conditions being fulfilled~\cite{MODULO}, 
the correlations from the hopping term can be removed and
the rotated model is finally known to be solvable by CBA.
\\
For  open boundary conditions 
the correlated hopping can be ``gauged away'' completely, yielding the HM 
without any boundary phases. 
\\
Instead, PBC lead to the HM with twisted boundary conditions.
Periodicity implies that the parameters $\xi_{i,j}$, $\alpha$, $A$ and 
$\gamma$ are periodic in their site-indices with period $L$.
The boundary phase is determined by hopping from site $L$ to 
site $L+1\,\hat{=}\,1$; 
such a phase enters as written in 
Eqs.~(\ref{BZERO1})--(\ref{BZERO2})
for the relations $\xi_{1,m}\leftrightarrow\xi_{L,m}$ and  
$\zeta_{1,m}\leftrightarrow\zeta_{L,m}$ and then for 
$A_{1,m}\leftrightarrow A_{L,m}$, $\alpha_{1,m}\leftrightarrow\alpha_{L,m}$.
Equations (\ref{ZERO1})--(\ref{BZERO2}) lead to
\begin{eqnarray}
2(\xi^{\sigma,-\sigma}_{1,m}-\xi^{\sigma,-\sigma}_{L,m}) &=&
\sum_{j=1}^{L-1}\alpha_{j,m}(\sigma) \stackrel{!}{=}\nonumber \\ 
&&\quad \stackrel{!}{=} -\alpha_{L,m}(\sigma)+ \phi^{(1)}_{\uparrow\downarrow}(\sigma),
\label{phase2}\\
2(\xi^{\sigma,\sigma}_{1,m}-\xi^{\sigma,\sigma}_{L,m}) &=&
\sum_{j=1 \atop j\neq m,m-1}^{L-1}A_{j,m}(\sigma) + \nonumber \\
&& \quad +2(\xi^{\sigma,\sigma}_{m-1,m}-\xi^{\sigma,\sigma}_{m,m+1})\nonumber\\ 
&&\quad\stackrel{!}{=} -A_{L,m}(\sigma)+ \phi^{(1)}_{\uparrow\uparrow}(\sigma), 
\label{phase1}\\
\zeta_{1,\sigma}-\zeta_{L,\sigma} &=&
\sum_{j=1}^{L-1}(\gamma_j(\sigma)+A_{j,j}) \nonumber \\
&\stackrel{!}{=}& -\gamma_{L}(\sigma)-A_{L,L}+\phi(\sigma), \label{phase0} 
\end{eqnarray}
where $\phi$ denotes the boundary phases. They can be determined 
(without solving  for the $\xi$'s explicitly) from Eqs. (\ref{phase2}) to (\ref{phase0}) as
\begin{eqnarray}
\phi^{(1)}_{\uparrow\downarrow}(\sigma)
&=& \sum_{j=1}^{L}\alpha_{j,m}(\sigma), \label{P1}\\
 \phi^{(1)}_{\uparrow\uparrow}(\sigma)
&=&\sum_{{j=1\atop j\neq m-1,m}}^{L} A_{j,m}(\sigma)+\nonumber \\ 
&& \quad +A_{m,m-1}(\sigma)+A_{m-1,m+1}(\sigma),\label{P2}\\
 \phi(\sigma)
&=&\sum_{j=1}^{L}\left (\gamma_j(\sigma)\;+\, A_{j,j}(\sigma)\right ).\label{scalarPhase}
\end{eqnarray}
Eqs. (\ref{INT1}) and (\ref{INT2})
also ensure that the phases in Eqs. (\ref{P1}) and (\ref{P2}) are 
$m$-independent\cite{NOTEPHASES}.
Summarizing, {\em iff}\/ Eqs. (\ref{INT1}) and (\ref{INT2}) are fulfilled,
the Hamiltonian~(\ref{SS}) can be mapped by $U$ (Eq. (\ref{U})) onto the  
usual HM with modified boundary conditions. The boundary twists are given by
\begin{equation}
\Phi_\sigma:=\phi(\sigma)+ \phi^{(1)}_{\uparrow\downarrow}(\sigma) N_{-\sigma} +
\phi^{(1)}_{\uparrow\uparrow}(\sigma)(N_\sigma-1) .
\end{equation}
We emphasize that here, the factor $(N_\sigma-1)$ appears 
instead of $N_\sigma$. The reason is that
the recursive relation (\ref{ZERO2}) does not exist for $m=j,j+1$.
But since one particle with spin $\sigma$ has to be on site $j$ and 
none on site $j+1$ (or vice versa: 
see the corresponding hopping term in~(\ref{SS})), one particle less accounts 
for the phase.

\subsection{Translational invariant models}

\begin{figure}[h]
\begin{minipage}{80mm}
\begin{minipage}[h]{40mm}
\psfig{width=40mm,height=40mm,file=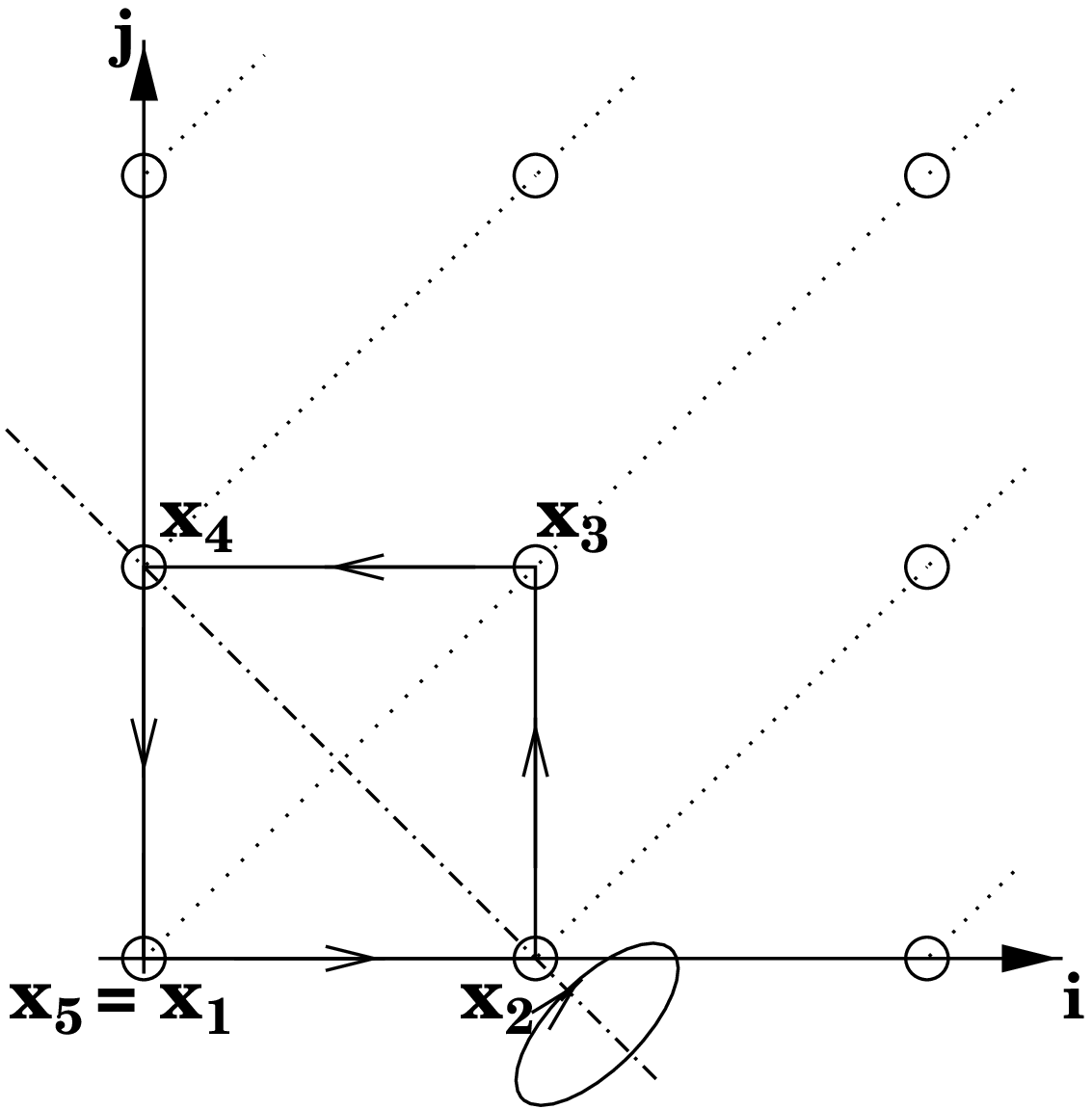}
\end{minipage}
\hspace{2mm}
\begin{minipage}[h]{32mm}
\psfig{width=32mm,height=22mm,file=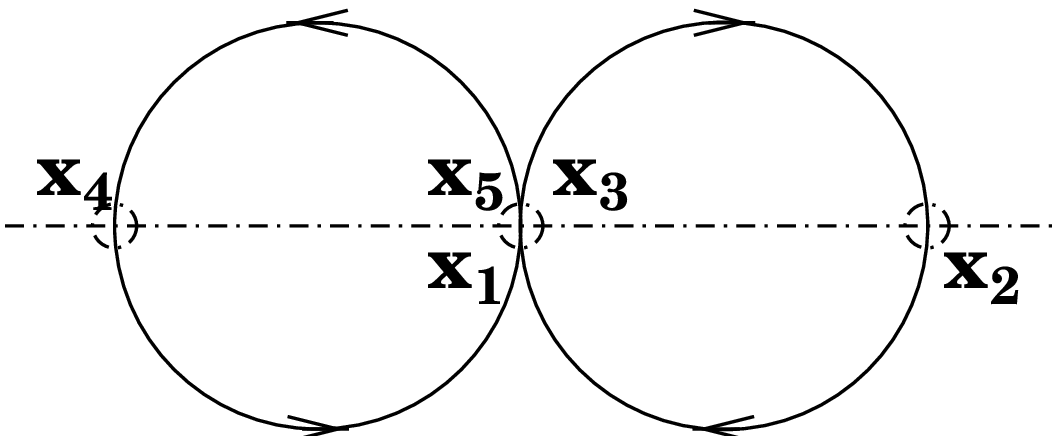}
\end{minipage}
\begin{center}
\begin{minipage}[h]{78mm}
\caption{\label{loops}
On the left, an elementary loop in two-coordinate space is drawn.
The circles are the coordinate sites, say $(i,j)$, corresponding to a parameter $\xi_{i,j}$.
If the parameters in the transformation are assumed
depending on the difference of the coordinates only, the dotted lines connect identified
sites in index space, corresponding to identical parameter values. 
One can imagine the resulting reduced lattice by furling the coordinate space
along the dash-dotted line to a cylinder or by pushing the identified sites on each other,
which is visualized on the right. In this case, the former elementary loop
turns into a double loop. Each of these two loops is an elementary loop.}
\end{minipage}
\end{center}
\end{minipage}
\end{figure}

In this section we assume  a translational invariant model, and hence translational
invariance of the parameters $\alpha$ and $A$:  
$\alpha_{j,l}(\sigma)=\alpha_{j-l}(\sigma)$,
$A_{j,l}(\sigma)=A_{j-l}(\sigma)$. 
\\
Restricting ourselves to translational invariant  
unitary transformations~\cite{NOTE} ($\xi_{j,k}=\xi_{j-k}$),
the former elementary loops become multiple loops (see Fig. \ref{loops}).
Then, the elementary loops are
$(j,m) \rightarrow (j+1,m) \rightarrow (j+1,m+1)$ and the closedness condition
only consists of the following terms from Eqs. (\ref{INT1}) and (\ref{INT2})
$\alpha_{j}(\sigma) = - \alpha_{-(j+1)}(-\sigma)$ for all $j$,
$A_{j}(\sigma) = - A_{-(j+1)}(\sigma)$ for all $j\neq 0$.
Also here, for the parameter $A$ not all elementary loops are viable.
But the only exceptional loop emerges from the case corresponding to 
$m=j$ in Eq. (\ref{INT2}) (that is a square on the diagonal -- the fat 
dotted line in Fig. \ref{criticalloop}).
But as already mentioned, the (excluded) condition for $m=j$ 
in Eq. (\ref{INT2}) is trivially fulfilled.
This is confirmed by calculating the effect of the jump from $0$ to $2$
in the exceptional loop in Fig.~\ref{criticalloop}, which here 
constitutes an elementary loop (comparing with
Fig.~\ref{loops} it turns out that the points $0$ and $2$ are identified
since translational invariant parameters are assumed).
As a consequence, the feature of additional phases imported by
subrelevant parameters and hence non-viable loops are absent here.
\\
But we want to stress that even here Eqs. (\ref{INT1}), (\ref{INT2}) are 
necessary and sufficient conditions for solvability --
for periodic as well as for open chains. This is because a
translational invariant model need not necessarily be gauged away 
by a unitary transformation with translational invariant parameters.
\\
Now, we can connect to Ref. \onlinecite{SCHULZ}. 
We obtain the boundary phases in Ref. \onlinecite{SCHULZ} through the 
identifications
$\alpha_{j,m}\rightarrow\alpha_{j-m}$
and $\alpha_j(\sigma):=-\sigma\alpha_j$.
In addition, $A_{j,m}(\sigma)=\gamma_j(\sigma)\equiv 0$ and  
$\beta_{i,j}=\sigma/2\cdot\xi^{\sigma,-\sigma}_{i,j}$,
giving an antisymmetric $\beta_{i,j}$.
However, in order to guarantee closedness and hence solvability of
the recursive relations, 
$$
\alpha_{j-m}-\alpha_{j-m-1}= -(\alpha_{m-j}-\alpha_{m-j-1})
$$
has to hold.

A special type of such models
is studied in Ref.~\onlinecite{KUNDU} with the only non-zero
values $\alpha_{0}=\alpha_{-1}=\eta $, where $\eta$ is a real
parameter. The closedness conditions are all fulfilled and therefore
the correlation factors in the hopping can be gauged away giving
exactly the phases found in Ref.~\onlinecite{KUNDU}. 
However, we could not reproduce the results following the path suggested 
in Ref.~\onlinecite{KUNDU} (see the general discussion in appendix~\ref{SCONFIG}).

\section{Multi-chain models}\label{MC-models}

Multi-chain models can be treated analogously. The chain-variable
plays the role of the spin-variable now. 
The only difference is that in general the chain variable ranges 
in a set different from $\{-1/2,+1/2\}$.
Since the method we employ to diagonalize the Hamiltonian stays 
the same, we only state the results.
For multi-chain models we assume {\it intra-chain hopping only}.

At first, we deal with spin-less fermions and let the hopping-term
of~(\ref{SS}) take the form 
\begin{equation}
a_{c,j+1}^\dagger a^{}_{c,j} \exp({\rm i}\gamma_j(c))
\exp\bigl[{\rm i} A_{j,l}^{c,d}n_{d,l}\bigr] + {\rm h.c.},
\end{equation}
where $c$ and $d$ are chain indices, which run in $\{1,\dots ,C\}$ with 
$C$ being  the number of chains~\cite{NOTECOLOR}.
The unitary transformation will be  
\begin{equation}
U:=\exp\bigl[ {\rm i}(\xi^{\mu,\nu}_{l,m}n_{\mu,l}n_{\nu,m}+\zeta_{\mu,l}n_{\mu,l})
\bigr],
\label{MULTI-U}
\end{equation}
where $\mu,\;\nu$ are now site indices (instead of spin indices of Sec. II),
running  in $\{1,\dots ,C\}$ (instead of $\{-1/2,1/2\}$). 
In this case, the recursive relations are 
\begin{eqnarray}
2(\xi^{c,d}_{j,m}-\xi^{c,d}_{j+1,m}) 
&\stackrel{!}{=}& A_{j,m}^{c,d};\ c\neq d, \label{ZERO11}\\
2(\xi^{c,c}_{j,m}-\xi^{c,c}_{j+1,m}) 
&\stackrel{!}{=}& A_{j,m}^{c,c};\ m\neq j,j+1, \label{ZERO12}\\
(\zeta_{c,j}-\zeta_{c,j+1}+2\xi^{c,c}_{j,j+1})
&\stackrel{!}{=}& \gamma_{c,j}. \label{ZERO13}
\end{eqnarray}
The closedness conditions read
\begin{eqnarray}
A_{j,m+1}^{c,d}-A_{j,m}^{c,d} &=& A_{m,j+1}^{d,c}-A_{m,j}^{d,c} \ ;\
(c\neq d)\ \vee \ m \neq j\pm 1,\label{INT11}\\ 
A_{j}^{c,d} &=& -A_{-(j+1)}^{d,c} \ ;\  c\neq d \ \vee\ j \neq 0,-1 \label{D-INT11} 
\end{eqnarray}
for all $c,d,j,m$ respecting the noted exceptions.
Equation (\ref{D-INT11}) applies to the translational invariant ansatz.
The boundary phases finally become
\begin{equation}
\Phi_c = \phi_c + \sum_{d=1}^{C}\phi^{(1)}_c (d) (N_{d}-\delta_{c,d}),
\end{equation}
where
\begin{eqnarray}
\phi^{(1)}_c (d)&=&\sum_{j=1}^L A^{c,d}_{j,m};\ c\neq d ,\label{Ph11} \\
\phi^{(1)}_c (c)&=&\sum_{{j=1\atop j\neq m-1,m}}^L \!\!\!A^{c,c}_{j,m}\
+ A^{c,c}_{m,m-1} + A^{c,c}_{m-1,m+1}, \label{Ph12}\\
\phi_c &=& \sum_{j=1}^L \left ( \gamma_{c,j} + A_{j,j}^{c,c} \right ).
\label{Ph13}
\end{eqnarray}
To compare this with Ref. \onlinecite{REPLY}, one has to replace 
$\xi^{\mu,\nu}_{l,m}$ with $\frac{1}{2}\xi_{\mu,\nu}B_{\mu,\nu}(l-m)$ and additionally
$A_{j,m}^{c,d}$ with $-\xi_{c,d}A_{c,d}(j-m)$.
There, $\xi_{\mu,\nu}$ was antisymmetric and hence, $B_{\mu,\nu}(l-m)$ has to be 
antisymmetric, too:
$B_{\mu,\nu}(m)=-B_{\nu,\mu}(-m)$.
The solvability conditions (\ref{D-INT11}) then transport into
\begin{neqblock}{c}
\xi_{c,d} A_{c,d}(j-m) = -\xi_{d,c} A_{d,c}(m-j-1) \Longleftrightarrow \quad \\ \\
A_{c,d}(j-m)= A_{d,c}(m-j-1)
\end{neqblock}
for all $m,j,c,d$.
This is equivalent to $A_{c,d}(-m)=A_{d,c}(m-1)$ for all $m,c,d$ as 
obtained in Ref. \onlinecite{REPLY}.
This condition is sufficient for CBA solvability but not necessary.
Necessary and sufficient is $A_{c,d}(-m)-A_{d,c}(-m-1)-A_{c,d}(m-1)+A_{d,c}(m)=0$
for all $m$.

If spin has to be included, nothing changes but the number of indices.
The hopping term becomes
\begin{equation}
a_{c,j+1;\sigma}^\dagger a_{c,j;\sigma} \exp({\rm i}\gamma_j(c))
\exp\bigl[{\rm i} A_{j,l}^{c,d}(\sigma,\rho)n_{d,l;\rho}\bigr] + {\rm h.c.},
\end{equation}
where $c,d$ are the chain indices again and $\sigma,\rho$ are spin indices.
The unitary transformation takes the form
\begin{equation}
U:=\exp\bigl[ {\rm i}(\xi^{\mu,\nu}_{l,m}(\rho,\tau)n_{\mu,l;\rho}n_{\nu,m;\tau}+
\zeta_{\mu,l;\rho}n_{\mu,l;\rho})
\bigr],
\end{equation}
where $\mu,\;\nu$ are the chain indices and $\rho,\tau$ are spin indices.
The recursion relations read
\begin{eqnarray}
2(\xi^{c,d}_{j,m}(\sigma,\rho)-\xi^{c,d}_{j+1,m}(\sigma,\rho)) 
&\stackrel{!}{=}& A_{j,m}^{c,d}(\sigma,\rho); \label{ZERO21}\\
 c\neq d\ \vee \ \sigma\neq\rho\ \vee\ m&\neq& j,j+1, \nonumber \\
(\zeta_{c,j}(\sigma)-\zeta_{c,j+1}(\sigma)+2\xi^{c,c}_{j,j+1}(\sigma,\sigma))
&\stackrel{!}{=}& \gamma_{c,j}(\sigma), \label{ZERO22}
\end{eqnarray}
and the closedness conditions are 
\begin{eqnarray}
A_{j,m+1}^{c,d}(\sigma,\rho)-A_{j,m}^{c,d}(\sigma,\rho) &=&
A_{m,j+1}^{d,c}(\rho,\sigma)-A_{m,j}^{d,c}(\rho,\sigma) ; \label{INT21} \\
 c\neq d\ \vee \ \sigma\neq\rho\ \vee\ m&\neq& j\pm 1, \nonumber 
\end{eqnarray}
for all $c,d,j,m,\sigma,\rho$ respecting the exceptions~\cite{NOTE-MULTI-INDEX}. 
The boundary phases become
\begin{equation}
\Phi_{c,\sigma} = \phi_{c,\sigma} + \sum_{d=1}^{C}
\sum_{\rho\in\{\uparrow,\downarrow\}}\phi^{(1)}_{c,\sigma}(d,\rho)
(N_{d,\rho}-\delta_{c,d}\delta_{\rho,\sigma}),
\end{equation}
where
\begin{eqnarray}
\phi^{(1)}_{c,\sigma} (d,\rho)&=&\sum_{j=1}^L A^{c,d}_{j,m}(\sigma,\rho);
\ c\neq d\ \vee\ \sigma\neq\rho \label{Ph21}\\
\phi^{(1)}_{c,\sigma} (c,\sigma)&=&\sum_{{j=1\atop j\neq m-1,m}}^L 
A^{c,c}_{j,m}(\sigma,\sigma)+\nonumber \\
&&\quad + A^{c,c}_{m,m-1}(\sigma,\sigma)+A^{c,c}_{m-1,m+1}(\sigma,\sigma), \label{Ph22}\\
\phi_{c,\sigma} &=& \sum_{j=1}^L \left ( \gamma_{c,j}(\sigma) + A_{j,j}^{c,c}(\sigma,\sigma) 
\right ).\label{Ph23}
\end{eqnarray}
The overlap of Ref. ~\onlinecite{SCHULZ} 
with the work presented in the Refs. \onlinecite{BOROVIK,ZVYAGIN} was 
discussed by the authors and was argued to not exist\cite{REPLY,COMMENT}. 
Now it is shown that the correlation in Refs. \onlinecite{BOROVIK,ZVYAGIN} 
cannot be gauged away, since the correlation parameters violate Eq. (\ref{INT21}).
However, the Hamiltonian having the slightly but essentially different phases
\begin{eqblock}{l}{ZVY-Correct}
\exp\Bigl\{-{\rm i}\pi\varphi_\sigma\bigl[
n_{l+1,m,\sigma}-n_{l-1,m,\sigma}+\\
\qquad\qquad + n_{l+1,m+1,\sigma}-n_{l-1,m+1,\sigma}
\bigr]+\displaystyle\frac{2\pi\gamma_\sigma}{N_a}\Bigr\}
\end{eqblock}
instead of
\begin{equation}
\exp\Bigl\{-{\rm i}\pi\varphi_\sigma\bigl[
n_{l+1,m,\sigma}-n_{l-1,m,\sigma}
\bigr]+\frac{2\pi\gamma_\sigma}{N_a}\Bigr\}
\end{equation}
yields the BE found in Reference\onlinecite{ZVYAGIN}.

The models studied in Ref.~\onlinecite{KUNDU2} are equivalent to multi-chain
models of spin-less fermions. In the first model, the only non-zero
model parameters are 
$A_{j,j}^{m,m+1}=-A_{j,j+1}^{m+1,m}=-4\Theta_{m,m+1}$ and 
$\gamma_j(m)=2(\Theta_{m+1,m}-\Theta_{m,m-1})$, where we used the
notation in Ref.~\onlinecite{KUNDU2}: $j$ and $m$ are the site and chain index
respectively.
The second model is represented by the parameters
$A_{j,j+1}^{m+1,m}=-A_{j,j}^{m,m+1}=\Theta_{m,m+1}+\alpha_{m,m+1}$ and
$A_{j,j}^{m+1,m}=-A_{j,j+1}^{m,m+1}=\Theta_{m,m+1}-\alpha_{m,m+1}$.
The closedness conditions are fulfilled 
{\em iff} $\Theta_{m,m+1} = \Theta_{m+1,m}$ and
$\alpha_{m,m+1}=\alpha_{m+1,m}$.

\section{Two-particle and higher correlated hopping}\label{2CH}

The procedure developed in the previous section can be extended to consider   
higher (than one-particle) correlated hopping.
First, we face explicitly 2--particle correlated hopping (2--CH). 
Then we will sketch how to deal with the general case of $n$--CH. 

2--CH corresponds to the occurrence of a term like 
$\bar{\alpha}_{l,m}^{\lambda,\mu}(j,\sigma) n_{l,\lambda} n_{m,\mu}$ in the exponential
factor of the hopping term of~(\ref{SS}). 
The 2-CH Hubbard-type Hamiltonian is
\begin{eqnarray}
\label{2-CH}
H &=& - t\, \sum_{j,\sigma } 
\, \biggl\{ c_{j+1,\sigma}^\dagger c_{j,\sigma} \exp({\rm i}\gamma_j(\sigma))\times \nonumber \\
&&\ \times\exp\left [ {\rm i}
{\bar{\alpha}}_{l,m}^{\lambda,\mu}(j,\sigma)n_{l,\lambda}n_{m,\mu}\right ]\times\\
&&\quad \times\exp\left [ {\rm i}{\alpha}_m^\mu (j,\sigma)n_{m,\mu}\right] + {\rm h.c.} 
\biggr\} + V\, \sum_{ i}n_{{i},\uparrow} n_{{ i},\downarrow}. \nonumber  
\end{eqnarray}
Without loss of generality the parameters 
${\bar{\alpha}}_{l,m}^{\lambda,\mu}(j,\sigma)$ 
can be chosen symmetric in  the index pairs 
$(l,\lambda)$ and $(m,\mu)$ and vanishing  if these index pairs 
coincide~(see Ref. ~\onlinecite{NOTEBLOCKS}). 
\\
The parameters ${\bar{\alpha}}_{j+1,m}^{\sigma,\mu}(j,\sigma)$ and 
${\alpha}_{j+1}^\sigma (j,\sigma)$ are irrelevant for all $j,m,\sigma,\mu$;
the effect of the subrelevant parameters on the 
lower correlated ones will be discussed later on in the present section.
\\
We first  remove the phases in the hopping term 
of (\ref{2-CH}) by a unitary transformation.
Then, we diagonalize the transformed Hamiltonian by CBA in computing
the boundary phases.
The 2--CH demands an exponent
$\bar{\xi}^{\lambda,\mu,\rho}_{l,m,r} n_{l,\lambda} n_{m,\mu} n_{n,\rho}$ in the unitary 
transformation $U$:
\begin{eqnarray}
U&:=&\exp\bigl[ {\rm i}(\bar{\xi}^{\lambda,\mu,\rho}_{l,m,r}n_{l,\lambda}n_{m,\mu}n_{r,\rho}+ 
\nonumber\\
&&\qquad + \xi^{\lambda,\mu}_{l,m}n_{l,\lambda}n_{m,\mu} + \zeta_{m,\mu}n_{m,\mu})
\bigr].
\end{eqnarray}
Both $\xi$ and $\bar{\xi}$ are totally symmetric and vanish if any two
pairs of parameters coincide.
\\
The hopping term is transformed into
\begin{eqnarray*}
c_{j+1,\sigma}^\dagger c_{j,\sigma} &\stackrel{U}{\longrightarrow}&
c_{j+1,\sigma}^\dagger c_{j,\sigma}\times \\
&&\ \times\exp\Bigl[3{\rm i}\bigl(\bar{\xi}^{\sigma,\lambda,\mu}_{j+1,l,m}-
     \bar{\xi}^{\sigma,\lambda,\mu}_{j,l,m}\bigr) n_{l,\lambda}n_{m,\mu} +\\
&& \quad + 2{\rm i} \bigl(\xi^{\sigma,\mu}_{j+1,m}-
     \xi^{\sigma,\mu}_{j,m} - 3\bar{\xi}^{\sigma,\sigma,\mu}_{j,j+1,m}\bigr) n_{m,\mu} +\\
&&\qquad + {\rm i}\bigl(\zeta_{j+1,\sigma}-\zeta_{j,\sigma}-2\xi^{\sigma,\sigma}_{j,j+1}
\bigr)
\Bigr],
\end{eqnarray*}
whereas the Coulomb interaction term remains unchanged.
This leads to the recursive relations (compare with~(\ref{ZERO1})--~(\ref{ZERO2}))
\begin{eqnarray}
3(\bar{\xi}^{\sigma,\lambda,\mu}_{j,l,m}-
  \bar{\xi}^{\sigma,\lambda,\mu}_{j+1,l,m})
&\stackrel{!}{=}& {\bar{\alpha}}_{l,m}^{\lambda,\mu}(j,\sigma), \label{2PC-ZERO1}\\
( m\neq l\ \vee\ \mu\neq\lambda)\ \mbox{and}\ 
(l,m\neq j,j+1 &\vee & \lambda,\mu\neq\sigma) \nonumber\\ 
2(\xi^{\sigma,\mu}_{j,m}-\xi^{\sigma,\mu}_{j+1,m} + 
3\bar{\xi}^{\sigma,\sigma,\mu}_{j,j+1,m}) 
&\stackrel{!}{=}& {\alpha}_m^\mu (j,\sigma); \label{2PC-ZERO2}\\
m\neq j,j+1\ \vee\ \mu &\neq&\sigma \nonumber\\
(\zeta_{j,\sigma}-\zeta_{j+1,\sigma}+2\xi^{\sigma,\sigma}_{j,j+1})
&\stackrel{!}{=}& \gamma_{j}(\sigma)\label{2PC-ZERO3}
\end{eqnarray}
for the parameters in $U$.
\\
We point out that, in the present case, 
two kinds of elementary loops exist because of the variety of indices 
in ${\bar{\alpha}}_{l,m}^{\lambda,\mu}(j,\sigma)$. Namely:
$(j;l,m)\rightarrow (j+1;l,m)\rightarrow (j+1;l+1,m)\rightarrow (j;l+1,m)
\rightarrow (j;l,m)$ and 
$(j;l,m)\rightarrow (j+1;l,m)\rightarrow (j+1;l,m+1)\rightarrow (j;l,m+1)
\rightarrow (j;l,m)$.
However, due to the symmetry of the $\bar{\alpha}$ both loops give the same closedness
condition for $\bar{\alpha}$, which is
\begin{equation}
\bar{\alpha}^{\lambda,\mu}_{l+1,m}(j,\sigma)-\bar{\alpha}^{\lambda,\mu}_{l,m}(j,\sigma)
=
\bar{\alpha}^{\sigma,\mu}_{j+1,m}(l,\lambda)-\bar{\alpha}^{\sigma,\mu}_{j,m}(l,\lambda)
\label{2PC-INT1}
\end{equation}
for $l\neq j,j\pm 1\ \vee\ \lambda\neq\sigma$ and 
$m\neq j,j+1(l,l+1)\ \vee\ \mu\neq\lambda$.\\
The corresponding boundary phases are
\begin{eqnarray}
\phi^{(2)}_\sigma (\lambda,\mu)&=&\sum_{j=1}^L 
{\bar{\alpha}}_{l,m}^{\lambda,\mu}(j,\sigma);\ \lambda,\mu\neq\sigma,\label{2Ph1}\\
\phi^{(2)}_\sigma (\sigma,\mu)&=&\sum_{{j=1\atop j\neq l,l-1}}^L 
{\bar{\alpha}}_{l,m}^{\sigma,\mu}(j,\sigma)+\nonumber\\
&&\ +{\bar{\alpha}}_{l-1,m}^{\sigma,\mu}(l,\sigma)+
{\bar{\alpha}}_{l+1,m}^{\sigma,\mu}(l-1,\sigma);\label{2P1'}\\
&&\mu\neq\sigma,\nonumber\\
\phi^{(2)}_\sigma (\sigma,\sigma)&=&\!\!\!\sum_{{j=1\atop j\neq l,l\pm 1}}^L 
\!\!\! {\bar{\alpha}}_{l,l+1}^{\sigma,\sigma}(j,\sigma)+\nonumber\\
&&\ +{\bar{\alpha}}_{l-1,l}^{\sigma,\sigma}(l+1,\sigma)+
{\bar{\alpha}}_{l-1,l+2}^{\sigma,\sigma}(l,\sigma)+\nonumber\\
&&\qquad\qquad\quad +{\bar{\alpha}}_{l+1,l+2}^{\sigma,\sigma}(l-1,\sigma).\label{2P1'''}
\end{eqnarray}
where $l\neq m, m\pm 1$ can be chosen arbitrarily. The 
result turns out to be independent of this choice. 
For less than three sites~\cite{NOTE-4SITES}, Eq. (\ref{2P1'''}) is ill-defined.
It reflects a physical limitation:
for the boundary phase  $\phi^{(2)}_\sigma (\sigma,\sigma)$
to occur, at least three particles with the same spin orientation have to exist; 
this is possible only if at least three sites are available. 
\\
Now we will discuss the effect of the subrelevant parameters 
$\bar{\alpha}^{\sigma,\mu}_{j,m}(j,\sigma)$ and $\alpha^\sigma_j(j,\sigma)$.
The 2-CH subrelevant part of the exponent in the hopping term is 
$2({\bar{\alpha}}_{j,m}^{\sigma,\mu}(j,\sigma)+
3{\bar{\xi}}^{\sigma,\sigma,\mu}_{j,j+1,m})$.
As discussed in the previous section, this term does not vanish in general\cite{SUBRELXI}
because the recursive relations do not cover the index grid of $\bar{\xi}$ completely.
It contributes to the 1-CH part instead.
It has to be added to the right-hand side of Eq. (\ref{2PC-ZERO2}).
As a consequence, the parameter 
$\bar{\xi}$ drops out and the recursive relation reads 
\begin{equation}
2(\xi^{\sigma,\mu}_{j,m}-\xi^{\sigma,\mu}_{j+1,m}) 
\stackrel{!}{=} {\beta}_m^\mu (j,\sigma) \ ;\ \ m\neq j,j+1 \ \vee\
\mu \neq \sigma ,
\label{2PC-ZERO2'}
\end{equation}
where $\beta^\mu_m(j,\sigma)$ is defined as  
\begin{equation}
\beta^\mu_m(j,\sigma):=\alpha^\mu_m(j,\sigma)+
2{\bar{\alpha}}_{j,m}^{\sigma,\mu}(j,\sigma).\label{2Predef1}
\end{equation}
Doing the same for the 1-CH subrelevant part in the hopping
(concerning Eqs. (\ref{2PC-ZERO3}) and (\ref{2PC-ZERO2'}) now), $\zeta$
drops out and in Eq. (\ref{2PC-ZERO3}),  $\gamma_j(\sigma)$  
will be substituted by $\tilde{\gamma}_j(\sigma)$ 
\begin{eqnarray}
\tilde{\gamma}_j(\sigma)&:=&\gamma_j(\sigma)+\alpha^\sigma_j(j,\sigma)+
2{\bar{\alpha}}_{j,j}^{\sigma,\sigma}(j,\sigma)\nonumber\\
&=&\gamma_j(\sigma)+\alpha^\sigma_j(j,\sigma)\label{2Predef2}.
\end{eqnarray}
Using Eq. (\ref{2PC-ZERO2'}), the second set of closedness conditions are obtained
\begin{equation}
\beta^\mu_{m+1}(j,\sigma)-\beta^\mu_{m}(j,\sigma) =
\beta^\sigma_{j+1}(m,\mu)-\beta^\sigma_j(m,\mu) \label{2PC-INT2}
\end{equation}
and the boundary phases are
\begin{eqnarray}
\Phi_\sigma &=& \phi_\sigma + \sum_{\lambda}\phi^{(1)}_\sigma (\lambda) 
(N_\lambda-\delta_{\lambda,\sigma})+\nonumber\\
&&\quad +\sum_{\lambda,\mu}\phi^{(2)}_\sigma (\lambda,\mu) 
(N_\lambda-\delta_{\lambda,\sigma}) (N_\mu-\delta_{\mu,\sigma}) ,
\end{eqnarray}
where
\begin{eqnarray}
\phi^{(1)}_\sigma (\mu)&=&\sum_{j=1}^L 
{\beta}_m^\mu (j,\sigma);\ \mu\neq\sigma, \\
\phi^{(1)}_\sigma (\sigma)&=&\sum_{{j=1\atop j\neq m-1,m}}^L 
{\beta}_m^\sigma (j,\sigma)+\nonumber\\
&&\quad {\beta}_{m-1}^\sigma (m,\sigma)+{\beta}_{m+1}^\sigma (m-1,\sigma), \\
\phi_\sigma &=& \sum_{j=1}^L \tilde{\gamma}_{j}(\sigma).
\end{eqnarray}
The analysis presented above can be generalized  to consider $n$-CH. 
\\
Let $\stackrel{(n)}{\alpha},\stackrel{(n)}{\beta}$ be the
parameters of the  $n$-CH part of the hopping:
\begin{eqnarray}
{\stackrel{(n)}{\beta}}^{\{{\cal S}\}}_{\{{\cal I}\}}(j,\sigma)&:=&
\sum_{k=0}^{\infty} {n+k \choose k}
{\stackrel{(n+k)}{\alpha}}^{\sigma,\dots,\sigma,\{{\cal S}\}}_{j,\dots,j,\{{\cal I}\}}
(j,\sigma)\label{REDEFINE}\\
&=& {\stackrel{(n)}{\alpha}}^{\{{\cal S}\}}_{\{{\cal I}\}}
(j,\sigma)+ (n+1){\stackrel{(n+1)}{\alpha}}^{\sigma\{{\cal S}\}}_{j\{{\cal I}\}}
(j,\sigma),\label{REDEF}\\
{\stackrel{(n)}{\beta}}^{\sigma,\dots}_{j,\dots}(j,\sigma)&:=&
{\stackrel{(n)}{\beta}}^{\sigma,\sigma,\dots}_{j,j,\dots}(.,.) :=0.\label{IRREL}
\end{eqnarray}
(for multi-chain models the spin index is a multi index spin/chain). 
The dots ($\dots $) stand for an arbitrary series of indices except they
appear in between two equal indices ($\sigma\dots\sigma$).
In this case, the void is meant to be filled up with $k$ times $\sigma$.
$\{{\cal S}\}$ is a set of spin- or color indices, $\{{\cal I}\}$ a set
of coordinate indices.
The sum in Eq. (\ref{REDEFINE}) of course is finite. The variables are
confined since $CL-1$ is the highest possible correlation level
if $C$ is the number of inner degrees of freedom ($C=2$ for spin
$1/2$). Hence, $k\leq CL-1-n$ and
for ${\bar{n}}$-CH with ${\bar{n}}\leq CL-1$, the sum in
Eq. (\ref{REDEFINE}) already stops at $\bar{k}={\bar{n}}-n$.
Furthermore, as mentioned above, parameters with coinciding index pairs
can be assumed to be zero. This leads to Eq. (\ref{REDEF}). 
Eq. (\ref{IRREL}) only reminds us that all subrelevant parts are removed
if $\beta$ is used instead of $\alpha$.
 
The closedness conditions take the form
\begin{equation}
\beta^{\lambda,\dots}_{l+1,\dots}(j,\sigma)-\beta^{\lambda,\dots}_{l,\dots}(j,\sigma)
=
\beta^{\sigma,\dots}_{j+1,\dots}(l,\lambda)-\beta^{\sigma,\dots}_{j,\dots}(l,\lambda)
. \label{genINT} 
\end{equation}
The $n$-CH boundary phase is given by
\begin{equation}\label{easyPhi}
\Phi^{(n)}_\sigma=\Phi^{(n-1)}_\sigma+\sum_{\lambda,\dots}\phi^{(n)}_\sigma(\lambda,\dots)
(N_\lambda-\delta_{\lambda,\sigma})\cdot\dots\; ,
\end{equation}
where the phases $\phi^{(n)}$ are
\begin{eqnarray}
\phi^{(n)}_\sigma(\{{\cal S}\})&=&\sum_{j=1}^L \stackrel{(n)}{\beta}^
{\{{\cal S}\}}_{\{{\cal I}\}}(j,\sigma);
\label{easydelta} \\
&&\qquad \sigma\not\in  \{{\cal S}\}, \nonumber\\
\phi^{(n)}_\sigma(\sigma,\dots,\sigma,\{{\cal S'}\})&=&
\!\!\!\sum_{{{j=1\atop j\neq l_i,l_i-1} \atop i=1..k}}^L 
\!\!\!\stackrel{(n)}{\beta}^{\sigma,\dots,\sigma,\{{\cal S'}\}}_
{l_1,\dots,l_k,\{{\cal I'}\}}(j,\sigma)+\nonumber \\
+ \sum_{i=1}^k &\Bigl[&
\!\!\stackrel{(n)}{\beta}^{\sigma,\dots,\sigma,\{{\cal S'}\}}_
{\dots,l_i-1,\dots,\{{\cal I'}\}}(l_i,\sigma)+
\nonumber\\
&+&\stackrel{(n)}{\beta}^{\sigma,\dots,\sigma,\{{\cal S'}\}}_
{\dots,l_i+1,\dots,\{{\cal I'}\}}(l_i-1,\sigma)\Bigr];
\label{easydelta2}\\
&&\qquad \sigma\not\in  \{{\cal S'}\}.\nonumber
\end{eqnarray}
The pair of index sets $\{{\cal S}\},\ \left|\{{\cal S}\}\right|=n$ and 
$\{{\cal I}\},\ \left|\{{\cal I}\}\right|=n$, respectively
$\{{\cal S'}\},\ \left|\{{\cal S'}\}\right|=n-k$ and 
$\{{\cal I'}\},\ \left|\{{\cal I'}\}\right|=n-k$ 
must not have coinciding index pairs and
in Eq. (\ref{easydelta2}) no two $l_i$ must be identical or neighbored.
Thus, Eq.~(\ref{easydelta2}) holds for $L\ge 2k$.
This validity range can be maximally enlarged using the 
closedness conditions (see Eq. (\ref{genINT})) 
\begin{eqnarray}
\phi^{(n)}_\sigma(\sigma,\dots,\sigma,\{{\cal S'}\})&=&
\!\!\!\sum_{{j=1\atop j\not\in [l,l+k]}}^L 
\!\!\!\stackrel{(n)}{\beta}^{\sigma,\dots,\sigma,\{{\cal S'}\}}_
{l+1,\dots,l+k,\{{\cal I'}\}}(j,\sigma)+\nonumber\\
&+& \sum_{i=0}^{k}
\stackrel{(n)}{\beta}^{\sigma,\dots,\sigma,\{{\cal S'}\}}_
{\{{\cal L}\}_{i,k},\{{\cal I'}\}}(l+i,\sigma),\label{generaldelta}\\ 
\{{\cal L}\}_{i,k}=[l,l+k+1]&\setminus &\{l+i,l+i+1\},\nonumber\\
\sigma &\not\in &  \{{\cal S'}\}\nonumber.
\end{eqnarray}
This formula holds for $L\ge k+1$, which is a limit set by physics --
analogous to the $2$-CH case.
The result is $l$-independent.
We point out that $\exp [{\rm i} F(\{n\})]$ with $F(\{n\})$ being 
an arbitrary functional of number operators is not the most general 
unitary operator in Fock space. The most general is
$\exp [{\rm i} G(\{c^\dagger, c\})]$ where
$G(\{c^\dagger, c\})$ constitutes a Hermitean functional of the complete set of
creation/annihilation operators.
In the Appendix~\ref{GENERAL-UNITARY} we will show
that the class of unitary operators discussed so far is 
large enough as far as removals of phases in a nearest neighbor
hopping term are concerned.

\section{Summary and conclusions}\label{CONCL}
In summary, we found a complete characterization of Coordinate Bethe Ansatz 
(CBA) solvable
Hubbard-type Hamiltonians with unitary correlated hopping for fermions.
A necessary and sufficient criterion for such a Hamiltonian
being solvable by CBA was formulated 
(see Eqs.~(\ref{INT1}), (\ref{INT2}), and (\ref{2PC-INT1})).
\\
In contrast to what is suggested in Ref.~\onlinecite{KUNDU}, we find that these
models are not CBA solvable in the ordinary plane waves basis. 
Indeed, in such a basis the scattering matrix is configuration
dependent (see Appendix~\ref{SCONFIG}) thus describing  diffractive
scattering. The particles interact non--trivially even for vanishing Coulomb
interaction. 
Such a situation resembles the case of the impurity
problem in the sense that also there, the free picture already
contains some residual interaction due to the impurity. 
Solvability of the models is recovered if the correlations from the
hopping terms can be gauged away. 
This is equivalent to equipping the plane-wave basis with additional 
density dependent phases. Only in this modified basis can the 
correlated hopping be absorbed in a boundary term. 
For the models considered in Refs.~\onlinecite{BOROVIK,ZVYAGIN} no such
basis exists and hence, they are not solvable by Bethe ansatz as they
stand. They can however be repaired by modifying slightly their
hopping term, as done at the end of chapter~\ref{MC-models}.
\\
The boundary twists for solvable models with periodic boundary
conditions are given explicitly in this work.
The corresponding Bethe equations are known from
Ref. \onlinecite{SUTHERLAND} adapting the boundary phases for a spin
degree of freedom only
\begin{eqnarray}\label{BETHEEQ}
&&{\rm e}^{{\rm i} p_j L} = {\rm e}^{-{\rm i}\Phi_\uparrow}
                \prod_{a=1}^{N_\downarrow}
                \frac{i(\sin p_j - \zeta_a) -\frac{V}{4t}}
                        {i(\sin p_j - \zeta_a) +\frac{V}{4t}}, \\  
&&\prod_{b=1 \atop b\neq a}^{N_\downarrow}
\frac{i(\zeta_a - \zeta_b) +\frac{V}{2t}}{i(\zeta_a - \zeta_b) -\frac{V}{2t}}
=  {\rm e}^{-{\rm i}(\Phi_\uparrow-\Phi_\downarrow)} 
          \prod_{l=1}^N
                \frac{i(\sin p_l - \zeta_a) -\frac{V}{4t}}
                        {i(\sin p_l - \zeta_a) +\frac{V}{4t}}. \nonumber   
\end{eqnarray}
Therein, $p_j$ are the quasi-momenta in the plane waves used in the
Bethe ansatz (see Eq.~(\ref{CBA})), $\zeta_a$ are spin rapidities, $t$ and $V$ are the 
Hubbard model parameters and $\Phi_\sigma$ are the boundary phases,
which have been determined in this paper. 
The connection to Ref. \onlinecite{SUTHERLAND} shows that in the
solvable cases, the correlation in the hopping 
term of the Hamiltonian (see, for instance,~(\ref{SS})) is
equivalent to applying a magnetic flux to the system. However, such a
flux is generated by the particles themselves (in particular, it is
not an external magnetic flux). Ground state properties can be deduced
from those calculated in Refs. ~\onlinecite{SUTHERLAND,AMOSECK,OAE}.
Even for absent Coulomb interaction, the many particle energy is not a
sum of single particle energies. The effect in the energy density is
of first order in $1/L$ and in $|\Phi_\uparrow - \Phi_\downarrow |$ in
the thermodynamic limit.  
So one can argue that correlated hopping accounts for a non trivial
interaction between the particles even for vanishing Coulomb
interaction.  
\\
The results obtained here can be applied to models for particles with
deformed exchange statistics(DES). This is done via a mapping from DES
to CH models\cite{OAE3}, whereas special DES models have been discussed recently
using direct Bethe ansatz~\cite{AMOSECK,OAE}. 
The details will appear in a forthcoming paper.

\acknowledgements

Motivating and fruitful discussions with D. Braak, M. Dzierzawa, M. Rasetti, 
P. Schwab, and B.S. Shastry
are gratefully acknowledged besides the support through the {\em Graduiertenkolleg} 
``Nonlinear Problems in Analysis, Geometry, and Physics" (GRK 283),
financed by the German Science Foundation (DFG) and the State of
Bavaria; this work was also partly supported by the SFB 484.

\begin{appendix}

\section{Configuration dependent $S$ matrix from standard CBA in
  presence of correlated hopping}
\label{SCONFIG} 
In this appendix we explain 
why correlated hopping destroys solvability by {\em direct}
CBA for Hubbard- and XXZ-type models.
\\
The CBA is an ansatz for the wave function in a so-called 
fundamental region~\cite{SUTHERLAND-NOTES}, in which the interaction term 
doesn't contribute. This fundamental region has to exist at first.
For the Hubbard model, this is guaranteed by particle-hole symmetries.
This gives the energy in terms of the distinct quasi-momenta.
The wave function has to be defined uniquely on the
intersection lines of the fundamental regions, where for the Hubbard model
also the interaction enters. Both demands yield 
the $S$ matrix,
which represents the effect of the interaction, a scattering of two particles.  
For Bethe ansatz solvability, the interaction must not have an effect beyond 
permuting the quasi-momenta of the particles. This demands that the
$S$ matrix fulfills the Yang-Baxter equation. 
Now the boundary conditions, if compatible with
the $S$ matrix, lead to additional conditions, fixing the eigenfunctions 
constructed by Bethe ansatz up to normalization. 
As a consequence, any additional condition destroys Bethe ansatz solvability.
Every correlation in the hopping further restricts the parameters in the Bethe ansatz.

Let us assume a given CH Hubbard-type model can be 
transformed iso-spectrally to another 
CH Hubbard-type model, for which fundamental regions exist.
This should mean that neither interaction nor correlated hopping contributes.
All correlations already contribute, when no interaction is yet to be included.
The Bethe Ansatz will have the form
\begin{equation}\label{CBA}
\psi(x_1,\dots,x_N)=:\sum_{\pi\in S_N} A(\pi|\pi'){\rm e}^{{\rm i}\sum\limits_{k=1}^N x_{\pi'(k)}p_{\pi(k)}}
\end{equation}
where the permutation $\pi'$ is chosen such that an appropriate order
in $(x_1,\dots,x_N)$ is achieved. Here, the chosen order which defines 
the fundamental regions, will be: $x_1\leq\dots\leq x_N$.

\subsection{``Pinned'' correlations}
At first assume a correlation which appears if a particle sits at a special site.
Let particle number $j$ sit on this site. 
The effect of this is a shift in all momenta by the correlation
strength $\varphi$, except that momentum of the designated particle. This is seen by 
projecting the Schr\"odinger equation on the specified configuration.
The particle causing the correlation with the others feels no correlation
from the hopping term applied to it, since it will be transported to
that site by the hopping. 
Thus the resulting term for the energy is
\begin{equation}
E=-2t\sum_{l\neq j}^N \cos(p_{\pi(l)}+\varphi) -2t\cos(p_{\pi(j)}).
\end{equation}
In this equation, $\pi$ is the momentum permutation from the Bethe ansatz.
This energy is neither independent of the permutation as it ought to be, nor
does it coincide with the original energy formula $E=-2t\sum_j \cos(p_j)$.
Note that $\varphi$ could even depend on the spin 
orientation of the considered particles.

\subsection{Relative correlations}
Assuming the hopping term to be
\begin{equation}
c^\dagger_{j+1,\sigma}c^{}_{j,\sigma}\exp({\rm i}\varphi n_{j+\Delta,\mu}) 
+ {\rm h.c.},
\end{equation}
where $\sigma$ and $\mu$ are spin indices.
Further assume one single particle at coordinate $j$ to be affected by
this correlation. The corresponding condition from the Schr\"odinger
equation is 
\begin{eqblock}{l}{CHSE}
\psi(j-1,\sigma;j',\mu)\bigl({\rm e}^{{\rm i}\varphi
  n_{j-1+\Delta,\mu}}-1\bigr) \nonumber\\ 
\quad + \psi(j+1,\sigma;j',\mu)\bigl({\rm e}^{{\rm i}\varphi
  n_{j+\Delta,\mu}}-1\bigr) = 0, 
\end{eqblock}
where we omitted the spin index from the argument of the wavefunction.
Two cases can appear independently from each other:
\begin{itemize}
\item{a particle with spin orientation $\mu$ sits at site $j'=j-1+\Delta$}
\item{a particle with spin orientation $\mu$ sits at site $j'=j+\Delta$}
\end{itemize}
leading to $\psi(j-1,\sigma;j-1+\Delta,\mu)=0$ in the first case and
$\psi(j+1,\sigma;j+\Delta,\mu)=0$ in the second case. These
constraints appear in addition to the usual ``continuity'' condition
arising from $\psi(j,\sigma;j,\mu)=-\psi(j,\sigma;j,\mu)$. If
$\sigma=\mu$, at the most two of these three conditions can coincide
if $\Delta=0$ or $\Delta=1$, which however constitute a subrelevant
and irrelevant correlation, respectively. 
For $\mu=-\sigma$, one of the correlation terms then coincides with
the on-site Coulomb interaction. The other, however, is still remaining.
The only way out is to modify the Bethe ansatz slightly~\cite{BARIEV}. 
In Ref.~\onlinecite{BARIEV}, the fundamental regions are without
intersection for different kinds of particles (different spin
orientations). Consequently, for different kinds of particles the
``continuity'' condition is absent and hence substituted by {\em one}
equation coming e.g. from correlated hopping. 
The second equation has then to coincide with the interaction.
This indicates that the hopping term
\begin{equation}\label{GOODCORR}
c^\dagger_{j+1,\sigma}c^{}_{j,\sigma}\exp({\rm i}\varphi n_{j+\Delta,-\sigma}) 
+ {\rm h.c.} \ ;\ \Delta\in\{0,1\}
\end{equation}
is the only possible relevant correlated
hopping which is tolerable by direct Bethe ansatz for 
Hubbard-type Hamiltonians (for XXZ-type models no
relevant CH is treatable by direct Bethe ansatz). 
But then, the obtained $S$ matrix
still had to fulfill the Yang-Baxter equation.
In Ref. ~\onlinecite{BARIEV}, this kind of hopping term is studied. There, $\varphi$ is
purely imaginary and $\Delta(\sigma):=(\sigma+1)/2\in\{0,1\}$ was chosen.
This Hamiltonian is shown to be solvable by a modified Bethe ansatz in
the absence of Coulomb interaction, $V=0$. For $V\neq 0$, the resulting $S$ matrix
no longer fulfills the Yang-Baxter equation.

To summarize, except for a very special type of relative correlation, namely with 
particles at distance zero or one towards raising site number, 
one obtains independent additional sets of 
conditions which the wave function had to fulfill. This obstructs
the direct Bethe ansatz, since the $S$ matrix becomes configuration
dependent. It then no longer factorizes into two-particle $S$ matrices.
Regarding the exceptional CH mentioned above, a configuration independent
$S$ matrix is obtained by applying a variant of Bethe ansatz~\cite{BARIEV}.
But for finite Coulomb-interaction strength $V$, this $S$ matrix does not
fulfill the Yang-Baxter equation. As a consequence, no CH XXZ-type model
and no interacting CH Hubbard-type model is tractable by direct Bethe
ansatz for finite $V$.

\section{A note on subrelevant parameters}
\label{DETAILS}
In the following, we will discuss the contribution of the subrelevant parameters.
Recalling the definitions, irrelevant parameters like $A_{j-1,j}(\sigma)$
do not at all affect the physics of the model, whereas subrelevant parameters
like $A_{j,j}(\sigma)$ contribute to the uncorrelated part of the hopping.
For this reason no recursive relations come from them.
It is worthwhile noting that 
Eq.~(\ref{INT2}) is trivially fulfilled for $\mbox{$m=j$}$.
For this case, the exponent produced after transformation by $U$ is
$i n_m(2\xi^{\sigma,\sigma}_{j+1,j}+A_{j,j}(\sigma))$.
This term hence appears as a phase additional to $\gamma_j(\sigma)$.
The parameters $\xi^{\sigma,\sigma}_{j+1,j}$ are (up to an additive constant)
given by jumping from $0$ to $2$ in the exceptional loop in figure~\ref{criticalloop}
\begin{equation}\label{DefXi}
2(\xi^{\sigma,\sigma}_{j,j-1}-\xi^{\sigma,\sigma}_{j+1,j})=
        A_{j,j-1}(\sigma)+A_{j-1,j+1}(\sigma).
\end{equation}
We will now consider the special case in which
$2\xi^{\sigma,\sigma}_{j+1,j}+A_{j,j}(\sigma)=0$ holds.
This is exactly what the relation (\ref{ZERO2}) would result in for $m=j$.
It implies that the subrelevant parameters do not create additional phases in the
uncorrelated part of the hopping.
Together with (\ref{DefXi}) it gives
\begin{equation}\label{NOPHASE}
A_{j,j}(\sigma)-A_{j,j-1}(\sigma)-A_{j-1,j+1}(\sigma)-A_{j-1,j-1}(\sigma)=0.
\end{equation}
The condition for the relations (\ref{ZERO2}) for $m=j$ and $m=j+1$ 
being consistent is $\mbox{$A_{j-1,j}=-A_{j-1,j-1}$}$, which can
always be fulfilled by
properly choosing the irrelevant parameter $A_{j-1,j}$.
Inserting this into Eq. (\ref{NOPHASE}), one finally also exactly obtains 
relation (\ref{ZERO2}) for $m=j+1$.
We can therefore conclude that if Eq. (\ref{NOPHASE}) holds, the contributions from
subrelevant parameters cancel out. It bridges the void in the recursive
relations (\ref{ZERO2}).
\\
Since all the $\alpha$'s are relevant, no voids occur in their recursive relations,
which is equivalent to all elementary loops being viable.

\section{Unitary transformations in the fermion $u(r)$ algebra}
\label{GENERAL-UNITARY}

In this appendix, the transformed
number operators $c^\dagger_i c^{}_i$ and hopping operators $c^\dagger_i c^{}_j$ 
will be studied for the most simple unitary transformation such as
\begin{equation}
U=\exp \left [ {\rm i}\alpha (c^\dagger_k c_l + c^\dagger_l c_k) \right ]\ ;\ k\neq l .
\end{equation}
The $r^2$ operators $\{c^\dagger_k c_l\; , \; 1\le i,j\le r \}$ span the Lie 
algebra $u(r)$:
\begin{equation}
[c^\dagger_i c_j, c^\dagger_k c_l]=\delta_{j,k} c^\dagger_i c_j- 
\delta_{i,l} c^\dagger_k c_j
\label{U(R)}
\end{equation}
where the Cartan basis is generated by $H_i=c^\dagger_i c_i$, 
$\left( i=1\dots r \right)$:
\begin{equation}
[H_i, c^\dagger_j c_k]= (\delta_{i,j}-\delta_{i,k}) c^\dagger_j c_k 
\end{equation}

The $u(r)$ algebraic structure allows us to write 
\begin{eqblock}{l}{EVIL}
{\rm e}^{{\rm i}\alpha\Phi_{k,l}^+ } n_m e^{- i\alpha\Phi^{}_{k,l}} = \\ \\
n_m+i\Phi_{k,l}^-(\delta_{l,m}-\delta_{k,m})- \\
\quad - \sin ^2\alpha (\delta_{l,m}-\delta_{k,m})(n_l-n_k)+ \\
\qquad\ + i(\sin\alpha\cos\alpha-\alpha)(\delta_{l,m}-\delta_{k,m})\Phi_{k,l}^- \\ \\
e^{ i\alpha\Phi_{k,l}^+} \Phi_{m,n}^+ 
e^{- i\alpha\Phi^{}_{k,l}} = \\ \\
\quad\Phi_{m,n}^+ 
+i\sin\alpha\bigl[\Phi_{k,n}^-\delta_{l,m}+\Phi_{l,n}^-\delta_{k,m}+ \\
\qquad\qquad  +\Phi_{k,m}^-\delta_{l,n}+\Phi_{l,m}^-\delta_{k,n}\bigr] +\\ \\
\quad +4\sin\alpha \bigl[\Phi_{k,l}^+(\delta_{k,m}\delta_{l,n}+\delta_{l,m}\delta_{k,n})+\\
\qquad\quad   +\Phi_{m,n}^+(\delta_{k,m}+\delta_{l,n}+\delta_{l,m}+\delta_{k,n})\bigr]
\end{eqblock}
where we applied the following equivalence for adjoints
$$
[Ad(\exp(A))]B:= e^{A} B e^{-A} = \sum_{n=1}^\infty
\frac{1}{n!}{[A,B]}_n =: \exp{[ad(A)]}B ,
$$ 
$${[A,B]}_{0}:=B\; ,\ {[A,B]}_{n+1}:=[A,{[A,B]}_n ],$$ 
and we have defined $\Phi^+_{k,l}:=c^\dagger_k c^{}_l+c^\dagger_l c^{}_k$, and
$\Phi^-_{k,l}:=c^\dagger_k c^{}_l-c^\dagger_l c^{}_k$
(the $\Phi^+$ are Hermitean, whereas the $\Phi^-$ are anti Hermitean).
\\
>From this it is seen that the transformation of the Hubbard Hamiltonian
creates arbitrary-range hopping from and to the sites $k$ and $l$, as well
as pair-hopping created from the interaction term. 
This can be understood from interpreting $\Phi_{k,l}^+$
as a Hamiltonian itself and $\alpha$ as the time. Then, Eq. (\ref{EVIL}) gives
the number and hopping operators in the Heisenberg picture.
>From this interpretation it seems reasonable that no linear combination
of $\Phi_{k,l}^+$ with $k\neq l$ will ever be able to just remove phases
in a nearest-neighbor hopping term. However, this is not absolutely true.
To point out the exceptional cases, the investigation has to be completed.
Since the identity $[Ad(\exp(A))]B=\exp{[ad(A)]}B$
does not considerably simplify calculating the action of a more general
unitary transformation, another approach will be taken.
But at first, the problem will be reduced as far as possible.

In a general product of creation and annihilation operators,
one can at first collect operators occurring in pairs using the exchange algebra.
The result is a multilinear form of number operators 
besides a multi-linear form of creation and annihilation operators, 
so that no two operators have coinciding indices.
So the most general exponent appearing in $U$ can be written as follows
\begin{eqblock}{l}{GENERALTRAFO}
\xi_{\{p\},\{q\}}\prod_i c^\dagger_{p_i} c^{}_{q_i} +
\xi_{k;\{p\},\{q\}}n_k \prod_i c^\dagger_{p_i} c^{}_{q_i} + \\ \\
\quad + \xi_{k,l;\{p\},\{q\}}n_k n_l\prod_i c^\dagger_{p_i} c^{}_{q_i} + \dots .
\end{eqblock}
Here, the different $\xi$ are symmetric in the indices before the semicolon and Hermitean
in the index sets behind it. They vanish if any two indices coincide.   
Assuming a pure $m$-linear form of  the creation/annihilation operators, then a
transformed bilinear object contains $m$-linear and even higher terms. 
This leads to multi-particle hopping
and interaction terms including more than two number operators \cite{EXTERN}.
Since the aim is to stay in the class of Hubbard- or XXZ-type Hamiltonians,
we consequently can limit ourselves to general multilinear forms of number
operators only, as already studied above, or bilinear forms of
creation/annihilation operators only.
We will discuss the latter in the following.
\begin{equation}\label{MIXEDTRAFO}
U:=\exp \bigl[ {\rm i} \xi_{k,l} c^\dagger_k c^{}_l \bigr]\ ;\ \xi^{}_{k,l}=\xi^*_{l,k}\; ,\; \xi_{k,k}=0 .
\end{equation}
We point out that $U c^\dagger_m c_n U^{-1}$ is bilinear,
since $[c^\dagger_k c_l,c^\dagger_m c_n]$ is bilinear~(see \ref{U(R)}). So, one can determine the result by projecting on the
desired initial and final states.
\begin{neqblock}{l}
\bra{0}c^{}_{m_f} \left [\xi_{k,l} c^\dagger_k c^{}_l\right]^n c^\dagger_{m_i}\ket{0} = \\
=\bra{0}c^{}_{m_f} \left [\xi_{k,l} c^\dagger_k c^{}_l\right] 
c^\dagger_{m_{n-1}}\ket{0}\bra{0}c^{}_{m_{n-1}}\dots  \\
\quad\dots c^\dagger_{m_1}\ket{0}\bra{0}c^{}_{m_1}\left [\xi_{k,l} c^\dagger_k c^{}_l\right]c^\dagger_{m_i}\ket{0} =\\
=\xi_{m_f,m_{n-1}}\xi_{m_{n-1},m_{n-2}}\cdot\dots\cdot \xi_{m_1,m_i} = {\left (\xi^n\right)}_{m_f,m_i},
\end{neqblock}
all written in sum convention. One directly obtains from this
\begin{eqnarray}
\bra{0}c^{}_{m_f} U c^\dagger_{m_i}\ket{0} &=& \left (\exp {\rm i} \xi\right )_{m_f,m_i} \label{UMatrix}\\
\bra{0}c^{}_{m_f} U c^\dagger_k c_l U^\dagger c^\dagger_{m_i}\ket{0} &=& 
\left (\exp {\rm i} \xi\right )_{m_f,k} \left (\exp {\rm i} \xi\right )^*_{m_i,l} ,\label{HopMatrix}
\end{eqnarray}
where hermitecity of the $\xi$ was used. The indices $k,l$ are not summed over.
It can be shown that the transformed interaction term can never include
a single-particle hopping term \cite{EXTERN}.
This already proves that the reverse direction is also impossible. 
Thus, number operators and nearest-neighbor hopping have to remain ``type-invariant'' 
under the transformation, since the type of Hamiltonian should be preserved. 
This results in restrictions on the matrix $\xi$
\begin{eqnarray}
\left (\exp {\rm i} \xi \right )_{m_f,k} \left (\exp {\rm i} \xi\right )^*_{m_i,k}&=&0
\ \forall\ m_f\neq m_i \label{CONDI}\\
\left (\exp {\rm i} \xi \right )_{m_f,j+1} \left (\exp {\rm i} \xi\right )^*_{m_i,j}&=&0
\ {\mbox for}\  |m_f-m_i|>1 .\label{CONDII}
\end{eqnarray}
The first conditions emerge from transforming the interaction term, whereas the second one
comes from the hopping term. Using both, we can deduce the structure
of $\exp{\rm i}\xi$.
\begin{neqblock}{l}
(\ref{CONDI})\ \Longrightarrow\ \forall\ k\ \exists_1 \ m(k)\ \ni\ 
        \left (\exp {\rm i} \xi \right )_{m(k),k}\neq 0 \\
(\ref{CONDII})\ \Longrightarrow\ m(k) = k + const. \\ \\
\Longrightarrow\ \left (\exp {\rm i} \xi \right )_{k,l}(\sDelta)=:
\delta_{l,k+\sDelta}r_k {\rm e}^{{\rm i}\phi_k} \ ; \ \prod_k r_l {\rm e}^{{\rm i}\phi_l} = 1.
\end{neqblock}
Unitarity of the matrix implies $r_l^2=1$ for all $l$. Hence it can be assumed that 
$r_l=1$ for all $l$.
So, one finally concludes that $U(\sDelta)$ transforms $c^\dagger_{j+1} c^{}_j$ into
$c^\dagger_{j+\sDelta+1} c^{}_{j+\sDelta} {\rm e}^{{\rm i}(\phi_{j+1}-\phi_j)}$. Number
operators remain unchanged. These phases can be gauged away leaving no
boundary phase. It is worth noting that pure $d$-range hopping
on an $L$-site chain can also be obtained {\it iff} $d$ and $L$ are relatively prime.
With the analysis used here it can finally be shown that unitary transformations
with multinomials of odd degree always produce particle-number violating terms.
Hence, they also make us leave the class of models we consider.
But it is clear from this, that a huge class of models, which is far from being 
Hubbard-type can be constructed by unitarily transforming the Hubbard model. 
They all are solvable and have the same spectrum as the Hubbard model.

\section{Proofs}\label{QED}
\subsection{Proof that transformations such as (\ref{GENERALTRAFO}) exceed the 
considered class of models}
\label{EXTERN}

Consider unitary transformations $U$ given by
\nbeqb{l}
U=\exp {\rm i}(\Omega +\Xi) \\ \\
\Xi= \xi_{\{p\},\{q\}}\prod_i c^\dagger_{p_i} c^{}_{q_i} + 
\xi_{k;\{p\},\{q\}}n_k \prod_i c^\dagger_{p_i} c^{}_{q_i} +\\ \\ 
\qquad + \xi_{k,l;\{p\},\{q\}}n_k n_l\prod_i c^\dagger_{p_i} c^{}_{q_i} + \dots ,
\neeqb
where $\Omega$ is a multilinear form of number operators only or
bilinear in creation/annihilation operators. 
Thus, the Hermitean $\Xi$ only contains terms higher than bilinear.

Now let us assume
\beq\label{ASSUME}
\exp( {\rm i}\Xi) n_k n_{k+1} \exp( -{\rm i}\Xi)= \sum_l a^k_l\cdot n_l n_{l+1}
\eeq
Where $a^k_l$ are constants for fixed $k$ and $l$.
This is impossible, because this expression contains terms at least
hexa-linear in creation/annihilation operators since $\Xi$ is at 
least quadri-linear. This still holds including $\Omega$.

Next assume 
\beq
\exp( {\rm i}\Xi) n_k n_{k+1} \exp( -{\rm i}\Xi)\propto
\exp ({\rm i} f[\{n\}]) c^\dagger_{l+1} c^{}_l
\eeq
and consider $\bra{0}c^{}_{m_f} \exp( {\rm i}\Xi) n_k n_{k+1}
\exp( -{\rm i}\Xi) c^\dagger_{m_i}\ket{0}$.
This had to result in $\delta_{m_f,l+1}\delta_{m_i,l}\exp ({\rm i}
g[\{n\}])$. Here, $f[\{n\}]$ is an arbitrary functional of the number
operators. In general, the functional $g[\{n\}]$ differed from $f[\{n\}]$.
However, $\bra{0}c^{}_{m_f} \Xi c^\dagger_{m_i}\ket{0} = 0$, since at
least two indices of 
each coefficient $\xi$ in $\Xi$ had to coincide in order to give a
non-zero contribution.  
But then the $\xi$ vanish themselves.
With the same argument, $\bra{0}c^{}_{m_f} \Xi^n
c^\dagger_{m_i}\ket{0} = 0$  for all integer $n$
(including zero, since $m_i\neq m_f$ is assumed).
Thus, the assumption cannot hold.
This completes the proof, since including $\Omega$ in the transformation $U$, 
the only nonvanishing contributions from 
$\bra{0}c^{}_{m_f} \exp {\rm i}(\Omega+\Xi) n_k n_{k+1}  
\exp -{\rm i}(\Omega + \Xi) c^\dagger_{m_i}\ket{0}$ come from terms
not including $\Xi$ at all. This means, that the only contributing
terms come from $\Omega$ alone. 

\subsection{Proof that the transform (\relax\ref{MIXEDTRAFO}) 
cannot interchange hopping and interaction}
\label{NOMIX}

Let us assume, $U n_j n_{j+1} U^{-1}$ (in case of XXZ-type models) contained a term 
$c^\dagger_{k+1} c^{}_k$.
Applying Eq.(\ref{HopMatrix}), this means
\beqb{l}{*}
(\exp {\rm i} \xi)_{m_f,j} (\exp {\rm i} \xi)^*_{m_i,j}\times \\
\quad\times (\exp {\rm i} \xi)_{\tilde{m}_f,j+1} (\exp {\rm i} \xi)^*_{\tilde{m}_i,j+1}
c^\dagger_{m_f} c^{}_{m_i}c^\dagger_{\tilde{m}_f} c^{}_{\tilde{m}_i} 
\eeqb
contained
a term $c^\dagger_{k+1} c^{}_k$. Here, $j$ is fixed and all the $m$'s are summed over.
Eq.(\ref{*}) implies
\begin{enumerate}
\item{$\tilde{m}_i=k$, $m_f=k+1$ and $m_i=\tilde{m}_f=:m$, or}
\item{$m_i=k$, $m_f=k+1$ and $\tilde{m}_i=\tilde{m}_f=:m$.}
\end{enumerate}
{\bf Considering 1.}:
Eq.(\ref{*}) in this case becomes
\beqb{l}{*'}
(\exp {\rm i} \xi)_{k+1,j} (\exp {\rm i} \xi)^*_{m,j}\times \\
\quad\times (\exp {\rm i} \xi)_{m,j+1} (\exp {\rm i} \xi)^*_{k,j+1}
c^\dagger_{k+1} c^{}_{m}c^\dagger_{m} c^{}_{k} ,
\eeqb
which is proportional to
\nbeqb{l}
(\exp -{\rm i} \xi)_{j,m}(\exp {\rm i} \xi)_{m,j+1} 
 (\exp {\rm i} \xi)_{k+1,j} (\exp {\rm i} \xi)^*_{k,j+1} \times \\
\qquad\times (1-n_m) c^\dagger_{k+1}  c^{}_{k} .
\neeqb
Since $m$ is summed over, the $1$ in the braces vanishes. 
This is because $\exp (-{\rm i}\xi) \exp {\rm i}\xi =\id$.
Additionally, no two-particle hopping must occur.
This implies that for arbitrary $m\neq\bar{m}$
\beqb{l}{**}
(\exp {\rm i} \xi)_{m,j+1} (\exp {\rm i} \xi)^*_{\bar{m},j} \times \\
\qquad\times \underbrace{(\exp {\rm i} \xi)_{k+1,j} (\exp {\rm i} \xi)^*_{k,j+1}}_{E_1} \stackrel{!}{=} 0.
\eeqb
Vanishing $E_1$ means no occurrence of nearest neighbor hopping. 
Hence, $E_1$ is assumed to be nonzero.
Since (\ref{*'})$\not\equiv 0\ \Longrightarrow$ for every fixed $j$
one can find an $m$ so that 
$(\exp {\rm i}\xi)_{m,j}\neq 0$ and $(\exp {\rm i}\xi)_{m,j+1}\neq 0$.
Now assume the existence of a further $\bar{m}\neq m$ so that even
$(\exp {\rm i}\xi)_{\bar{m},j}\neq 0$. 
This implied $(\exp {\rm i}\xi)_{m,j+1}(\exp {\rm i}\xi)_{\bar{m},j}\neq 0$ in contradiction to Eq.(\ref{**}).
Thus, there exists only a single $m$ for which $(\exp {\rm i}\xi)_{m,j+1}$ and $(\exp {\rm i}\xi)_{m,j}$ don't vanish.
But this would mean $\det \exp {\rm i}\xi = 0$, which would contradict
the assumptions.

{\bf Considering 2.}:
Eq.(\ref{*}) now becomes
\nbeqb{l}
(\exp {\rm i} \xi)_{k+1,j+1} (\exp {\rm i} \xi)^*_{k,j+1}
 (\exp - {\rm i} \xi)_{j,m} (\exp {\rm i} \xi)^*_{m,j} \times \\
\qquad\times c^\dagger_{k+1} c^{}_{k}c^\dagger_{m} c^{}_{m} .
\neeqb
As above, no two-particle hopping must occur.
This implies that for arbitrary $m\neq\bar{m}$
\beqb{l}{***}
\underbrace{(\exp {\rm i} \xi)_{k+1,j+1} (\exp {\rm i} \xi)^*_{k,j+1}}_{E_2}\times \\
\qquad \times (\exp {\rm i} \xi)_{\bar{m},j} (\exp {\rm i} \xi)^*_{m,j} \stackrel{!}{=} 0.
\eeqb
Vanishing $E_2$ means no occurrence of nearest neighbor hopping. Hence, 
$E_2$ is assumed to be nonzero.
Since $\det \exp {\rm i}\xi \neq 0$, one finds for every fixed $j$ an $m$ 
so that $(\exp {\rm i}\xi)_{m,j}\neq 0$.
Now assume  a further $\bar{m}\neq m$ to exist so that 
$(\exp {\rm i}\xi)_{\bar{m},j}\neq 0$. 
This implies $(\exp {\rm i}\xi)_{m,j}(\exp {\rm i}\xi)_{\bar{m},j}\neq 0$ in 
contradiction to Eq.(\ref{***}).
Thus, there exists only a single $m$ for which $(\exp {\rm i}\xi)_{m,j}$ is nonzero.
This is in contradiction to $E_2$ being nonzero.

This completes the proof.

\end{appendix}

\end{multicols}


\begin{thebibliography}{100}
\bibitem{NOTESS} For spin 1/2 every function of $f(\sigma)$ is linear in $\sigma$:
        $f(\sigma)=:f_0 + \sigma f_1$, where $f_0$ and $f_1$ are constant.
        In the Ref. ~\onlinecite{SCHULZ} only $\alpha(\sigma)$ exists, which there 
	was chosen to be a 
        periodic translational invariant function of the site indices. 
        Additionally, $\alpha(\sigma)=\sigma \alpha_1$ was taken.
\bibitem{MODULO} In fact the phases have to be removed modulo
        $2\pi$ only, which therefore also holds for the closedness
        conditions. However, having a parameter which scales the
        strength of the correlation (i.e. 
        $A^{\sigma,\mu,\dots}_{j,m,\dots}=\varphi 
        \tilde{A}^{\sigma,\mu,\dots}_{j,m,\dots}$), the equations
        are not to be taken modulo $2\pi$ as far as
        solvability for arbitrary values of $\varphi$ is concerned. 
        Also for this reason ${\rm mod}\; 2\pi$ is omitted throughout the paper.
\bibitem{NOTEPHASES} The phases (\ref{P1}) and (\ref{P2}) can also be obtained 
        from calculating the contribution
        of the exceptional loop in Fig. \ref{criticalloop} (respecting
        the formerly unknown 
        boundary phases). Setting the contribution of this loop to zero determines 
        the phase. In case of (\ref{P1}), no forbidden crossings
        exist, as mentioned before.
\bibitem{NOTE}In fact, the unitary transformation need not be translational
        invariant even for a translational invariant model.
        Such an ansatz only means looking for a solution in a subset of 
        unitary transformations.
        Hence, equations~(\ref{INT1})--(\ref{INT2}) are still necessary
        but no more sufficient for the closedness of the recursive relation
        for the translational invariant ansatz, but they are still for
        CBA solvability.
\bibitem{NOTECOLOR}The chain index can also be
        interpreted as a color index, corresponding to some other inner 
        degrees of freedom, characteristic for the particles.
        But all the models discussed here have in common that this
        index labels a good quantum number, corresponding to a conserved quantity.
        For models including color, color conservation is assumed.
\bibitem{NOTE-MULTI-INDEX}In fact, spin and chain variables could be combined 
        into a multi-index. Then the spin had the effect of doubling
        the number of chains.
\bibitem{NOTEBLOCKS} We point out that from now on we group the indices 
	into blocks.
        Namely all indices of a corresponding number operator belong to 
        one such block,
        containing a site- and a spin/color (multi-) index. 
        An index exchange always means an exchange of a whole index block.
\bibitem{NOTE-4SITES}Actually, an equation similar to Eq. (\ref{2P1'''})
        but valid for $L \ge 4$ only can be obtained by 
        extending Eq. (\ref{P2}) straight forwardly. Eq. (\ref{2P1'''})
        has already been 
        suitably extended to its maximum range of validity $L\ge 3$,
        applying the closedness condition Eq. (\ref{2PC-INT2}).
\bibitem{SUBRELXI} From the recursive relations, 
        $\bar{\xi}^{\sigma,\sigma,\dots}_{j,j+1,\dots}$ 
        expressed in terms of $\bar{\alpha}$ is
        \begin{eqnarray*}
          \bar{\xi}^{\sigma,\sigma,\mu}_{1,2,m}(j,\sigma)&:=& 0 \\
          3\bar{\xi}^{\sigma,\sigma,\mu}_{j,j+1,m}(j,\sigma)&=& - 
          \sum_{k=2}^{j}({\bar{\alpha}}_{k-1,m}^{\sigma,\mu}(k,\sigma)+  \\
          &&\qquad\quad +{\bar{\alpha}}_{k+1,m}^{\sigma,\mu}(k-1,\sigma));\\
          &&\qquad\qquad  j<L,\\
          3\bar{\xi}^{\sigma,\sigma,\mu}_{L,1,m}(j,\sigma)&=& - 
          \sum_{k=2}^{L}({\bar{\alpha}}_{k-1,m}^{\sigma,\mu}(k,\sigma)+ \\
          &&\qquad\qquad +{\bar{\alpha}}_{k+1,m}^{\sigma,\mu}(k-1,\sigma))+\\
          &&\qquad\qquad\quad - \phi^{(1)}_\sigma(\sigma,\mu).
        \end{eqnarray*}
        But since they drop out of the recursive equations, these 
        relations are not needed here.
\end{thebibliography}
\end{document}